\documentclass[prl,onecolumn,superscriptaddress]{revtex4}
\usepackage{amsmath,amssymb,mathrsfs}

\usepackage{hyperref}
\usepackage{paralist}
\usepackage{pdfpages}
\usepackage{lscape} %
 \usepackage{graphicx}

\def\be{\begin{equation}}
\def\ee{\end{equation}}
\def\bea{\begin{eqnarray}}
\def\eea{\end{eqnarray}}
\def\nn{\nonumber}

\newcommand{\Eq}[1]{Eq.~\eqref{#1}}

\newcommand{\beq}{\begin{equation}}
\newcommand{\eeq}{\end{equation}}

\newcommand{\beqa}{\begin{eqnarray}}
\newcommand{\eeqa}{\end{eqnarray}}

\newcommand{\Beqa}{\begin{eqnarray*}}
\newcommand{\Eeqa}{\end{eqnarray*}}

\newcommand{\I}{\mathrm{i}}

\begin{document}

\title{Momentum distribution and non-local high order correlation functions of  1D strongly interacting Bose gas}

\author{EJKP Nandani}
\affiliation{
Wuhan Institute of Physics and Mathematics, Chinese Academy of Sciences, Wuhan 430071, China.}
\affiliation{University of Chinese Academy of Sciences, Beijing 100049, China.}
\affiliation{Department of Mathematics,University of Ruhuna, Matara 81000, Sri Lanka.}



\author{Xi-Wen Guan}
\email[e-mail:]{xwe105@wipm.ac.cn}
\affiliation{
Wuhan Institute of Physics and Mathematics, Chinese Academy of Sciences, Wuhan 430071, China.}
\affiliation{Department of Theoretical Physics, Research School of Physics and Engineering,
Australian National University, Canberra ACT 0200, Australia.}

\date{\today}

\begin{abstract}
 \paragraph*{ Abstract :}
 The Lieb-Liniger model is a prototypical integrable model  and has  been turned into the  benchmark physics in theoretical and numerical investigations of low dimensional quantum systems.
 In this note, we present various methods for calculating local and nonlocal $M$-particle  correlation functions,  momentum distribution and static structure factor.
In particular,  using the Bethe ansatz wave function of the strong coupling Lieb-Liniger model,    we analytically calculate two-point correlation function, the large moment tail of momentum distribution and static structure factor of the model in terms of the fractional statistical parameter $\alpha =1-2/\gamma$, where $\gamma$ is the dimensionless interaction strength.
We also discuss the Tan's adiabatic  relation and other universal  relations for the strongly repulsive Lieb-Liniger model in term of the fractional statistical parameter.
 %


 \paragraph*{Keywords:} Correlation function, Momentum distributions, Structure Factor, Contact
\end{abstract}

\maketitle
%

\section{I. Introduction}

The Bethe ansatz, which was  introduced in $1931$ by Hans Bethe, has become  a powerful  method to obtain exact solutions of  one-dimensional (1D)  quantum many-body systems.
In 1963, Lieb and Liniger  ~\cite{Lieb} solved the 1D many-particle problem of $\delta$-function interacting bosons by
the Bethe's hypothesis.
The ground state, the momentum, and elementary excitations were obtained  for this model by using the Lieb-Liniger solution.
In this context, a significant step was made on the discovery of the grand canonical   description of this Lieb-Liniger model by Yang and Yang in $1969$~\cite{Yang}.
Now, this grand canonical approach  is  called Yang-Yang thermodynamic method.
The Yang-Yang thermodynamics of the  Lieb-Liniger Bose gas provides   benchmark   understanding of quantum statistics, thermodynamics and quantum critical phenomena in  many-body physics, see a review~\cite{Korepin-book,Jiang:2015}.
In the context of ultracold atoms, the  1D Bose gas with a repulsive short-range interaction characterized by a tunable  coupling constant exhibits rich many-body  properties.
This model  thus becomes an ideal test ground to explore fundamental many-body phenomena ranging from  equilibrium to  nonequilibrium physics in the experiment~\cite{Paredes:2004, Kinoshita:2004,Haller:2009,Yang:2017,Kinoshita:2006, Hofferberth:2007}.\\

 Despite Lieb-Liniger is arguably  the simplest integrable model,  the calculation of correlation functions is extremely challenging due to the complexity of the  Bethe ansatz many-body wave function of the model.
The study of  correlation functions has  long been being  an important theme in the physics of ultracold quantum gases since they  provide information of quantum many-particle interference and coherence  beyond the solely spectra of the systems.
Therefore,  there has been much  theoretical and experimental interest in   the local, non-local, and dynamical correlation functions at zero and finite temperatures via numerous methods based on exactly solvable models, see \cite{HCF,dmg_gvs,Kheruntsyan,dmg1,Toshiya,Kormos:2010,Kormos:2011,Astrakharchik1,Astrakharchik_2006,Caux_2009,Caux_2014,Caux_2006}.\\

For sufficiently strong  interaction and sufficiently low  density,  the 1D Lieb-Liniger  gas  enters the  Tonks-Girardeau (TG) regime, in which bosons behave like impenetrable particles (hard-core bosons).
Such  impenetrable bosons behave mostly like the free fermions  that  build up  the  Girardeau's Bose-Fermi  mapping~\cite{Girardeau}.
In fact,  the 1D Lieb-Liniger Bose gas with
the interacting strength $c_{\rm B}$ can map onto the fully-polarized  fermions with a p-wave interaction of strength  $c_{\rm F}=1/c_{\rm B}$  \cite{Cheon:1999}.
As a result of the Bose-Fermi mapping, the energy spectra of the Bose and corresponding Fermi system are identical at the TG regime.
The  observables that can be given in terms of the local density  are identical for  both systems, such as dynamical density-density correlation function, see ~\cite{Alexander}.
However, this mapping for the off-diagonal correlation function does not like to  be true, for example, momentum distribution.
 The Bose-Fermi mapping has tremendous applications in the study of strongly interacting quantum gases of ultracold atoms.
An expansion of the reduced density matrices for $1D$  bosons has been first derived by Lenard ~\cite{Lenard_1964,Lenard_1966} using bose-fermi mapping at Tonks Girardeau regime.  After that, there were superb  findings in the literature, for example, \cite{PJForrester_2003,PJForrester_2006,Jimbo,Raoul,HGVaidya,HGVaidya_1979,Caux_2007,Jimbo_1980}.\\

In the context of correlation functions, the Fourier coefficients have an important physical significance.
By taking Fourier transform of the one- and two-particle correlation functions, the momentum distribution and the static structure factor of  the system can be obtained, respectively.
The first few terms of the short-range series expansion of the one-particle correlation function
$g_{1}(x)=1+\sum_{i=0}^\infty c_i|nx|^i$
have been calculated in ~\cite{Caux_2007,Maxim_2003}, and ~\cite{Maxim1_2003}, where $c_1=0,\; c_2=1/2(e(\gamma)-\gamma e^\prime(\gamma))$ and $c_3=\gamma^2e^\prime(\gamma)/12$, etc.
A further study on the connection between the non-local one-body and local three-body correlation functions was presented in \cite{Olshanii:2017}.
%
%
The systems described by the Hamiltonian with short distance interaction ($\delta$-function type) show  singular behaviour of the wave function when two or more  colliding particles coincide.
Consequently,   the large momentum tail of the momentum distribution has a universal power-law decay, i.e. $w(p)\sim C/p^4$ where $C$ is an extensive quantity called Tan's contact \cite{Tan-coantact, Tan1:2008,Tan2:2008}.
Beside the power law tail of the momentum distribution, the
contact characterizes many universal properties of the
systems that are independent of the details of the contact  interaction and the dimensionality ~\cite{Rafael}, such as the energy adiabatic  relation, the universal  connection to the structure factor  and  the pressure relation.
The contact measuring the correlation of such systems of this kind   in short distance limit is  proportional to the
probability of finding two atoms at zero separation~\cite{Tan-coantact, Tan1:2008,Tan2:2008}, see more studies on the contact in \cite{zhang:2009,Werner:2009,Braaten:2019}.
The contact  can be computed from the free energy of the system via the Helmann-Feynmann theorem~\cite{Ovidiu_2017}, or from  the prefactor of the high-frequency power-law decay of the single particle correlator~\cite{Joaquin}, or by the operator product expansion field theory method~\cite{Barth:2011}, etc.
Very recently, p-wave contacts were found to exhibit universal relations in the ultracold atomic systems with a p-wave interaction~\cite{Stewart:2015,Yoshida:2015,Cui:2016}.
In this scenario, the momentum distribution,  s-wave and p-wave contact relations have been  attracted a great deal of attention from theory and experiment~\cite{Yuta,Maxim1_2003, Maxim_2003,Xu:2015,Ovidiu_2017,Minguzzi,Rafael,Wild_2012,Yin:2018}.\\

On the other hand, the static structure factor contains information about the static correlation properties  of a system and directly relates to the pair  distribution function or the normalized density-density correlator or spin-spin correlator~\cite{Alexander}.
In the low-momentum region and the infinite repulsive regime, hydrodynamic theory predicts that the static structure factor should behave linearly with momentum, $k$~\cite{Rafael}.
The static structure factor at small momentum $k$ is related to the sound velocity, i.e. $(v_s)$ by $S(k)=\hbar|k|/(2mv_s)$~\cite{Xiwen1,Astrakharchik1,Alexander,Caux_2014,Rafael,Caux_2006}.
For high momentum, the structure factor always converges to $1$~\cite{Rafael}.
The Fourier transform of the time-dependent density-density correlation function is known as the dynamical  structure factor~\cite{Alexander_2005,Caux_2006,Caux_2015}, and expresses the probability of excitation by an infinitesimal perturbation.\\

The  duality between interacting spinless bosons and  fermions in 1D ~\cite{Cheon:1999,Alexander,Raoul} can be adopted  to  the framework of the  generalized fractional statistics, which interpolate continuously between bosons and fermions.
 In fact, while three dimensions particles can be only bosons or fermions, in lower dimensionality it has been recognized that particles with fractional statistics intermediate between bosons and fermions~\cite{Wu:1994,Bernard:1994,Murthy:1994,Ha:1995,Anyon,guan:2006,Guan_2007} due to the transmission between dynamical interaction and statistical interaction.
The $1D$ strongly repulsive  Bose gas at low temperatures is
equivalent to  a  gas of ideal particles obeying the non-mutual generalized exclusion statistics (GES) with a statistical parameter $\alpha =1-2/\gamma$, where $\gamma$ is the dimensionless interaction strength \cite{Guan_2007}.
Using such a mapping,
i.e. the quasimomenta of $N$ strongly interacting bosons map to the momenta of $N$ free fermions via $k_i\approx \alpha k_i^F  $ with $i=1,\ldots, N$~\cite{Guan_2007}.
By using the fractional statistical parameter $\alpha$,  the $M$-body local correlation functions and non-local correlation functions in terms of the wave function of $M$ bosons at zero collision energy and zero total momentum have been computed analytically up to next-to-leading order ~\cite{HCF}.
In this paper, we analytically calculate various correlation functions in terms of the  statistical parameter $\alpha$.
In particular, we calculate the two-point correlation function,  the power-law tail of momentum distribution and discuss the structure factor
of $1D$ Lie-Liniger model of ultracold atoms  with finitely strong  interaction strength.
For Tonks gas, i.e.  $\alpha=1$,  our  results are equivalent to that of  the Tonks-Girardeau gas.\\

  The paper is organized as follows: In sections I and II, we briefly introduce  the Lieb-Liniger  model and   recall  some results of higher order local  correlations for the 1D interacting bosons.
 In section III,  we present the derivation of the nonlocal $2M$-point correlation function.
 In Section IV,  we prove the power-law decay of the  momentum distribution of the 1D interacting bosons at large momenta.
The large momentum tail of the momentum distribution determines Tan's contact in term of the fractional statistical parameter $\alpha$.
In this section,  we  also discuss the static structure factor.
The last  section  presents our conclusion. 

\section{I. Model of interest}

A model of $1D$ spinless $N$ bosons with $\delta$-function interaction  is described by  the Hamiltonian
\beq\label{H}
H_N = \dfrac{\hbar^2}{2m}\Big( \sum\limits_{j=1}^N-\partial^2_{x_j} + 2c\sum\limits_{1\leq j<i\leq N}\delta(x_j - x_i)\Big),
\eeq
where $m$ is the mass of bosons  and $c=-2/a_{1D}$  is a coupling constant.
Here, $a_{1D}$ is the 1D scattering length and $\hbar$ is the Plank constant. The positive and negative $c$ values correspond to the repulsive and attractive interactions, respectively, and $c=0$ for free particles. \\

With the periodic boundary condition, $\psi(0,x_2,\ldots,x_N)=\psi(x_2,\ldots,x_N,L),$ in the domain $0\leq x_1\leq x_2\leq\ldots\leq x_N\leq L$,  the Betha ansatz wave function  can be written as~\cite{Lieb}

 \beq\label{BA_wave}
 \psi(x_1,\ldots,x_N)=\sum\limits_P(-1)^P\Big[\prod\limits_{1\leq i< j\leq N}\Big(\dfrac{\I c+k_{P_i}-k_{P_j}}{\I c-k_{P_i}+k_{P_j}}\Big)^{1/2}\Big] e^{\I \sum\limits_{j=1}^Nk_{P_{j}}x_j},
 \eeq
 where $(-1)^P$ is parity for the permutations 
    and the sum runs over $N!$ permutations $P$ with respect to quasi-momentum $\{k_1,k_2,\ldots,k_N\}$.
    After expanding the amplitude of the wave function,  we have

   \beq\label{wave}
   \psi(x_1,\ldots,x_N) =\begin{cases}
   \mathcal{A}(c)\sum\limits_P(-1)^P  e^{\I \sum\limits_{j=1}^Nk_{P_{j}}x_j},&\text{for strong coupling},\\
   N!-\dfrac{c}{2}Y(x_1,\ldots,x_N),&\text{for weak coupling},
   \end{cases}
   \eeq
   where $\mathcal{A}(c)=\big(1+\frac{1}{c}\sum_{j=1}^{N}(2j-N-1)\partial_{x_j}\big)$ and $Y(x_1,\ldots,x_N)=\sum\limits_P\big(\sqrt{\frac{L}{2}}\sum\limits_{i<j}  F(q_{P_i}-q_{P_j})+\sqrt{\frac{2}{L}}\sum\limits_{j=1}^N\; q_{P_j}x_j\big)^2$,  here $F(q_{i}-q_{j})=(q_i-q_j)^{-1}$ and $q_j=k_j\sqrt{L/2c}$.
  The derivation of the wave function for weak coupling can be seen in  {\bf Appendix A}, also see the case in the 1D p-wave fermions in \cite{Yin:2018}.
   Moreover, the normalization factor of the wave function is defined by
$\mathcal{N}^2=\int_{0}^{ L} dx_1\ldots dx_N\; |\psi(x_{1},x_{2},\ldots,x_{N})|^2$.
 Further,  the wave function is continuous whenever two particles close to each other, and the discontinuity in the derivative of the wave function is $2c$ when two particles overlap. These are known as boundary conditions of the wave function with $\delta$ potential~\cite{Lieb}.
 For a given set of quasi-momenta$\{k_j\},$ the total momentum and the energy of the system are  $\mathcal{P}=\sum_{j=1}^N k_j$ and $E=\sum_{j=1}^N k_j^2$, respectively.\\

   For a  finite large  interaction strength, the quasimomenta  of the bosons $k_1, k_2, \ldots, k_N$ deviate from pure Fermi statistics,
 they obey the non-mutual general exclusive statistics (GES) ~\cite{Wu:1994,Bernard:1994,Guan_2007}.
 The deviation from Fermi statistics is determined  by the non-mutual  GES  parameter  $\alpha=1-2/\gamma$ ~\cite{Guan_2007}.
 If the total momentum $k_1+\cdots+k_N=0$, we have the following  relation:
 \beq\label{GSE}
 k_j=k_j^F\alpha+O(c^{-2}),
 \eeq
  where
 $k_j^F=2\pi \mathcal{I}_j/L$ and the $\mathcal{I}_j$'s are integers satisfying $\mathcal{I}_1<\mathcal{I}_2<\cdots<\mathcal{I}_N$.
In fact, the equal spacing in momenta reveals a  nature of non-mutual fractional statistics.
Here, the wave function of $1D$ interacting bosons  can be given in terms of  the GES $ k_j=k_j^F\alpha$, 
 \beq\label{GSE_x2}
 \psi(x_1,\ldots,x_N) =\begin{cases}
  \Big(1+\frac{\alpha}{c}\sum_{j=1}^{N}(2j-N-1)\partial_{x_j^F}\Big)\psi_N^F,& x_j=x_j^F/\alpha,\\
  \Big(1+(\alpha-1)\sum\limits_{j=1}^{N} x_j^F\partial_{x_j^F}+\frac{1}{c}\sum\limits_{j=1}^{N}(2j-N-1)\partial_{x_j^F}\Big)\psi_N^F,& x_j\sim x_j^F,
   \end{cases}
 \eeq
where  $\psi_N^F=\sum_P(-1)^P exp(\I\sum_{j=1}^N k_{P_j}^Fx_j^F)$ is the wave function of $N$ free fermions.

\section{II. Local $M$-body correlation functions}

We first would like to recall some results of local $M$-body correlation functions which  have been studied \cite{HCF,dmg_gvs,Kheruntsyan,dmg1,Toshiya,Kormos:2010,Kormos:2011,Astrakharchik1,Astrakharchik_2006,Caux_2009,Caux_2014,Caux_2006}. In particular, we discuss the results of the correlation functions  which were recently derived by using the GES relation \Eq{GSE}. It turns out that the GES is convenient to calculate the local $M$-body  correlation functions.

\subsubsection{II.1 ~~Strong coupling regime}

The local $M-$particle correlation is a measure of the probability of observing $M$ particles at the same position and defined by
\beqa\label{mbody}
g_M&=&\dfrac{N!}{(N-M)!}\dfrac{\int_0^L\ldots \int_0^L dx_{M+1}\ldots dx_N\;|\psi(0,\ldots,0,x_{M+1},\ldots,x_N)|^2}{\int_{0}^{ L}\ldots \int_0^L dx_1\ldots dx_N\;|\psi(x_1,\ldots,x_N)|^2}.
\eeqa
In the strong coupling limit,  one can  expand  the BA wave function, \Eq{wave} in terms of the power of $1/c$ from which the numerator and denominator can be calculated analytically \cite{HCF}
 \beq\label{gM}
   g_{M}=\frac{N!\,c^{-M(M-1)} \int_{0}^{L}\ldots \int_0^L\Big|\bigtriangleup_{M}(\partial_x)\phi(x_1,\cdots,x_N)|_{x_1=\cdots=x_M=0}\Big|^2dx_{M+1}\cdots dx_N+O(c^{-M(M-1)-2})}{(N-M)!\left[ 1+\frac{2N(N-1)}{c L}+O(c^{-2}) \right]\int_0^L\ldots \int_0^L\Big|\phi(x_1,\ldots,x_N)\Big|^2dx_1\cdots dx_N}.
 \eeq
Here,  $\Delta_{M}(\partial_x)=\prod\limits_{1\le i<j\le  M}\left(\frac{\partial}{\partial x_{j}}-\frac{\partial}{\partial x_{i}}\right)$ is the Vandermonde determinant and
$\phi(x_1,\ldots,x_N)=\sum\limits_P (-1)^Pe^{\I\sum_{j=1}^N k_{Pj}x_j}$, for all $x$'s.
In this case, using  the GES \Eq{GSE}, $k_i=k_i^F\alpha$ and making a scaling change  $x_i= x_i^F/\alpha $ for $i=1,\ldots, N$, thus we can write \Eq{gM} as
\beq
   g_{M}=\frac{N!}{(N-M)!}\frac{\frac{\alpha^{M^2-N}}{c^{M(M-1)}}\int_{0}^{L}\ldots \int_0^L\Big| \Delta_{M}(\partial_{x^F})\phi^F(x_1^F,\cdots,x_N^F)|_{x_1^F=\cdots=x_M^F=0}\Big|^2dx_{M+1}^F\cdots dx_N^F}
{\alpha^{-(N-1)}\int_0^L\ldots \int_0^L|\phi^F(x_1^F,\cdots,x_N^F)|^2dx_1^F\cdots dx_N^F},\label{GES-form}
\eeq
to the order $O(c^{-M(M-1)-2})$. Here, $ \phi^F(x_1^F,\cdots,x_M^F)\equiv\sum_P(-1)^Pe^{\I\sum_{j=1}^N k_{Pj}^Fx_j^F}$.
Since $\phi^F(x_1,\ldots,x_N)$ is a Slater determinant, by applying Wick's theorem
 and changing variable $k^F=2\pi n z$, then the $M$-body correlation function reads
\beq
g_M=M!\,n^{M}\big(2\pi/\gamma\big)^{M(M-1)}\alpha^{M^2-1}\int_{-\infty}^\infty \ldots \int_{-\infty}^{\infty} dz_1\cdots dz_M\,\Big(\prod\limits_{j=1}^{M}N(z_j)\Big)\Delta_M^2(z)
+O(\gamma^{-M(M-1)-2}),\label{G-M}
\eeq
where the dimensionless parameter  $\gamma=c/n$,  the liner density is given by $n=N/L$ and $N(z)=\left(1+exp\big(\frac{\frac{\hbar^2}{2m}(2\pi n z)^2-\mu}{k_BT}\big)\right)^{-1}$, $T$ and $\mu$ are  the temperature and the chemical potential of the 1D Bose gas.
Calculating the multiple integrals in \Eq{G-M}   by using the orthogonal polynomials in the random matrix  theory~\cite{Madan}, we can give the $M$-particle local correlation function as
\beq{
g_{M}=(M!)^2\,n^{M}\big(2\pi/\gamma\big)^{M(M-1)}\alpha^{M^2-1}\prod_{i=0}^{M-1}h_{i},}
\eeq
where $h_{i}$'s are the norm-squares of  the monic orthogonal polynomials satisfying
$\int_{-\infty}^{\infty}P_{i}(z)P_{j}(z)N(z)dz =h_i\delta_{ij}.$
Considering the  distribution of  particles at  ground state and applying Sommerfeld expansion,  we have  general formulas for the higher order local  correlation function in two temperature regimes \cite{HCF}
\beq\label{s2body}
\dfrac{g_{M}}{n^M}=\begin{cases}
\dfrac{\big(\prod_{j=1}^{M} j!\big)^2\alpha^{(M^{2}-1)}(\pi/\gamma)^{M(M-1)}}{\big(\prod_{j=1}^{M-1}(2j-1)!!\big)^2(2M-1)!!}\bigg\{1+ \frac{1}{48} M^2 (M^{2}-1)\left(\dfrac{T}{\pi T_d}\right)^{2}+O\Big[\Big(\frac{T}{T_d}\Big)^2\Big]\bigg\},
& T\ll T_{d},\\
(M!)^{2}\alpha^{M^2-1}\left(\dfrac{T}{2\gamma^2T_d}\right)^{\frac{M(M-1)}{2}}\prod\limits_{j=0}^{M-1}j!, &T_{d} \ll T\ll \gamma^{2}T_{d},
\end{cases}
\eeq
where $T_d\equiv\hbar^2n^2/(2mk_B)$ is the quantum degeneracy temperature.
In the above calculation, the statistical parameter $\alpha$  in the quasi-momenta of the hardcore  bosons can be extracted as a factor giving the next-order corrections  in the correlation function (\ref{s2body}).
In such a way, the integrant in the correlation function (\ref{GES-form}) reduces to the Vandermonde determinant with respect to the the wave function of the  free fermions.
Then   the  Wick's theorem and random matrix theory  are   used to obtain the explicit form of $M$-body correlation function (\ref{s2body}).
However, it would be extremely difficult to obtain  higher order correction terms beyond the form of (\ref{s2body}).

\subsubsection{ II.2~~Weak coupling regime}

For weak coupling case, one can easily use the wave function (\ref{wave}) to calculate the $M$-body correlation function
\beqa
&&g_M^w(y_1,\ldots,y_M;x_1,\ldots,x_M)\nn\\
&=&\dfrac{N!}{(N-M)!\mathcal{N}^2}\int_0^L\ldots \int_0^Ldx_{M+1}\ldots dx_N(N!-\dfrac{c}{2}Y(y_1,\ldots,y_M,x_{M+1}\ldots,x_N))^*(N!-\dfrac{c}{2}Y(x_1,\ldots,x_N))\nn\\
&=&\dfrac{N!}{(N-M)!\mathcal{N}^2}\Big[N!^2L^{N-M}-N!\dfrac{c}{2}\int_0^L\ldots \int_0^Ldx_{M+1}\ldots dx_N[(Y(y_1,\ldots,y_M,x_{M+1}\ldots,x_N))^*+Y(x_1,\ldots,x_N)]\Big]\nn\\
&&\hspace{15cm}+O(c^2).
\eeqa
So, the $M-$particl local-correlation function becomes
\beq
g_M^w(0,\ldots,0;0,\ldots,0)=\dfrac{N!}{(N-M)!\mathcal{N}^2}\Big[N!^2L^{N-M}-N!c\int_0^L\ldots \int_0^L dx_{M+1}\ldots dx_N[Y(0,\ldots,0,x_{M+1},\ldots,x_N)]\Big]+O(c^2),
\eeq
where $Y(0,\ldots,0,x_{M+1},\ldots,x_N)=\sum\limits_P\Big(\sqrt{\dfrac{L}{2}}\sum\limits_{i<j}\Big[\;F(q_{P_i}-q_{P_j})\Big]+\sqrt{\dfrac{2}{L}}\sum\limits_{j=M+1}^Nq_{P_j}x_j\Big)^2$.

In fact, the  Bogoliubov method is very convenient and useful   for the study of the weak coupling Bose gas in any dimension.
 In this Bogoliubov approach, one consider  a large fraction of particles in the lowest quantum state and excites a small fraction of particles out of the Bose-condensed particles with weakly repulsive interacting Bose gas.
 Therefore, in this regime, the field operator,$\hat\Psi$, can be considered as a sum of the condensate part, $\psi_{0}$, and a non-condensate part, $\psi^{\prime}$, such as
 $ \hat\Psi =\psi_{0}+\psi^{\prime}$.
 The density of the system is $ n=n_0+n^\prime$
where $n_0=\langle\psi_0\psi_0^*\rangle$  and  $n^\prime=\langle\psi^{\prime\dagger}\psi^\prime\rangle$ represent  the  condensed particles density and non condensed particles density, respectively.
For a uniform condensate, the ground state solution is $\psi_0=\sqrt{n_0}exp(\I\mu t/\hbar)$, where the chemical potential is given by the relation $\mu=n_0g$, see a detailed study  given in  \cite{dmg_gvs}.

In the weak coupling regime,$\gamma\ll 1$, the mean field interaction energy per particle proportional to $ng$, where  $g=\hbar^2 c/m$, and it is proper to define the correlation length, $l_c=\hbar/\sqrt{mng}$.
At temperature $T\ll T_d$, over  a wide range of parameters $l_c\ll l_\phi$, we can have the condition  $T/T_d\ll\sqrt{\gamma}$~\cite{dmg_gvs}.
Here, $l_\phi$ is the phase coherence length and $T_d=\hbar^2n^2/2m$ is the quantum degeneracy temperature.
In this case, the condensate fraction contains the contribution of excitations with momenta $k\leq k_0\ll l_c^{-1}$, and non-condensate parts provided excitations with $k\sim l_c^{-1}$~\cite{dmg1,dmg_gvs}.
 In order to complete our discussion on the local $M$-body correlation, the necessary calculation of the correlation function  in {\bf Appendix B}.
 What below is a  summary of the correlation functions based on the  Bogoliubov approach.

  The $M$-particle correlation function for $1D$ bosons at weak coupling regime is given by
 \beq
 g_M^w=\big<(\psi_0^{\ast}+\psi^{\prime\dagger})^M(\psi_0+\psi^{\prime})^M\big>
 = n^M\Big(1+\frac{M(M-1)}{n}\big(\langle\psi^{\prime\dagger}\psi^{\prime}\rangle+\langle\psi^\prime\psi^\prime\rangle\big)\Big),\label{eq}
 \eeq
dealing with the $M^{th}$ order density
$n^M= n^M-Mn^{M-1}\langle\psi^{\prime\dagger}\psi^{\prime}\rangle+O\big(\langle\psi^{\prime\dagger}\psi^\prime\rangle\big)^2$. The
 \Eq{eq} shows that $g_M^w$  depends on the normal density, $\langle\psi^{\prime\dagger}\psi^{\prime}\rangle$, and anomalous density, $\langle\psi^\prime\psi^{\prime}\rangle$, of the non-condensed part. Considering the form of the non-condensate part:
 $\psi^\prime=\sum\limits_k\Big(u_k \hat b_k e^{-\I\epsilon_k t/\hbar}-v_k^{\ast}\hat b_k^\dagger e^{\I\epsilon_k t/\hbar}\Big)$,
  $g_M^w$ can be converted into
   \beq
 g_M^w=n^M\Bigg(1+\frac{M(M-1)}{2\pi n}\int_{-\infty}^{\infty}dk\Big(\frac{E_k}{\epsilon_k}(N_k+1)-1\Big)\Bigg),\label{e3}
  \eeq
 by Bogoliubov transformation, where $\hat{b}_k,\hat{b}_k^\dagger$ are operators of excitations which obey the usual commutation relations, $\epsilon_k$ are their eigen energies, and $u_k,v_k$ their eigenfunctions.
   This result illustrates that $g_M^w$ depends on the occupation number of the excitation,$N_k$, excitation energy, $\epsilon_k$, and system energy $E_k$. In fact, $g_M^w$ varies with the temperature range since $N_k=0,\; \epsilon_k\approx\sqrt{E_k^2+2gn_0E_k}$ at zero temperature and $N_k\neq0,\;\epsilon_k=\hbar kv_s$ at non-zero temperature. Here, $v_s\approx\sqrt{ n_0g/m}$ is the sound velocity and the distribution $N_k$ takes the form  $(e^{\epsilon_{k}/ T}-1)^{-1}$  for the temperature $T\ll\mu$ and $T\gg\mu$, respectively.
 However, $E_k=\hbar^2k^2/2m$  and $\mu\approx n_0g$ at low  temperature.

This study followed the method developed by  Gangardt and  Shlyapnikov in~\cite{dmg_gvs}.
 The elementary excitation of the Bose-condensed system involves vacuum fluctuations and thermal fluctuations. At  zero temperature, thermal fluctuations are suppressed and vacuum fluctuations dominate.
 This scenario becomes reversed at $\mu\ll T\ll\sqrt{\gamma}T_d$.
However, both  vacuum and thermal fluctuations create excitations at $T\ll\mu$.

 According to these configurations, and after a length algebra,   the $M$-particle correlation, \Eq{e3}  is  given by
  \beq\label{weak_gm}
  \dfrac{g_M^w}{n^M} =\begin{cases}
 1-\dfrac{M(M-1)}{\pi }\sqrt{ \gamma },& T=0,\\
  1-\dfrac{M(M-1)}{\pi }\sqrt{ \gamma }+\dfrac{M(M-1)T^2\pi}{48} \dfrac{\sqrt{\gamma}}{(\gamma T_d)^2},&T\ll\mu,\\
  1+\dfrac{M(M-1)T }{4\pi\sqrt{\gamma} T_d },&\sqrt{\gamma}T_d\gg T\gg\mu.
  \end{cases}
  \eeq
 These results are not valid for $T\geq T_d\sqrt{\gamma}$, belonging to quantum decoherence regime. The detailed calculation is presented  in  {\bf Appendix B}.
This result  has been reported in \cite{dmg_gvs}.
The study $2$-body correlations was also  reported  in~\cite{Kheruntsyan}, while the   $2$-body, $3$-body correlation functions  at zero temperature were studied in~\cite{dmg1}.

\subsection{III.~Non-local correlation functions}

The  $M$-particle non-local correlation is a measure of the probability of observing $M$ particles at different points and defined by
\beqa\label{mbody} \small
&&g_M\left(x_1,\ldots,x_M;x_1^{'},\ldots,x_M^{'}\right)\nn\\
&=&\dfrac{N!}{(N-M)!}\dfrac{\int_0^L \ldots \int_0^L dx_{M+1}\ldots dx_N \psi^*(x_1^{'},\ldots,x_M^{'},x_{M+1},\ldots,x_N) \psi(x_1,\ldots,x_M,x_{M+1},\ldots,x_N) }{\int_{0}^{ L}\ldots \int_0^L dx_1\ldots dx_N\;|\psi(x_1,\ldots,x_N)|^2}\nonumber \\
&=&\dfrac{N!}{(N-M)!}\dfrac{\int_0^L\ldots \int_0^L dx_{M+1}\ldots dx_N \psi^*(0,\ldots,0,x_{M+1},\ldots,x_N) \psi(x_1-x_1^{'},\ldots,x_M-x_M^{'},x_{M+1},\ldots,x_N) }{\int_{0}^{ L} \ldots \int_0^L dx_1\ldots dx_N\;|\psi(x_1,\ldots,x_N)|^2}\nonumber\\
&=& g_M\left(x_1-x_1^{'},\ldots,x_M-x_M^{'};0,\ldots,0\right).
\eeqa
The Galilean invariance is used in  the above equation.
The short distance non-local  correlation functions
$\langle\Psi^\dagger(x_1^{'})\cdots\Psi^\dagger(x_M^{'})\Psi(x_M)\cdots\Psi(x_1)\rangle$
of the strongly repulsive Bose gas was studied  in terms of the wave function of $M$ bosons at zero collision energy and zero total momentum in \cite{HCF}.

\subsubsection{III.1~ Strong coupling regime}

We first consider the  non-local correlation function of strongly interacting $1D$ bosons  with $c=\infty$.
Let's shorten our notation
$g_{M}(x_1,\ldots,x_M)=g_M\left(x_1,\ldots,x_M;0,\ldots,0\right)$.
 According to the Girardeau's Bose-Fermi mapping~\cite{Girardeau},
 we have
 $g_{M}^B(x_1,\ldots,x_M)\approx |g_{M}^F(x_1,\ldots,x_M)|$.
 Thus the $M$-particle correlation is rewritten as
\beqa\label{nl-gM}
 g_{M}(x_1,\ldots,x_M)=\dfrac{N!}{(N-M)!}\dfrac{ \int_{0}^{ L}\ldots \int_0^L dx^F_{M+1}\ldots dx_N^F\big|\psi_N^F\big|^2}{\int_{0}^{ L} \ldots \int_0^L dx^F_{1}\ldots dx^F_N\big|\psi_N^F\big|^2},
 \eeqa
where $N$-particle fermion wave function, $\psi_N^F$ has a determinant form and norm of the wave function can be expressed as $\big|\psi^F_N\big|^2=\det\Big(\dfrac{\sin(N\pi(x^F_j-x^F_l)/L)}{\sin(\pi(x^F_j-x^F_l)/L)}\Big)\Big|_{j,l=1,\ldots,N}$ and the normalization factor  $\int_{0}^{ L} \ldots \int_0^L dx^F_{1}\ldots dx^F_N |\psi_N^F|^2=N!L^N$. By changing the variable as $t_j=2\pi x^F_j/L$,  \Eq{nl-gM} becomes
 \beqa\label{gt}
g_{M}(t_1,\ldots,t_M)&=&\dfrac{(2\pi)^M}{(N-M)!L^M}\int_{0}^{ 2\pi} \ldots \int_0^{2\pi} dt_{M+1}\ldots dt_N
 \Big|\det\Big(\dfrac{\sin(N(t_j-t_l)/2)}{2\pi\sin((t_j-t_l)/2)}\Big)\Big|_{j,l=1,\ldots,N}.
\eeqa

{\em Theorem:}~\cite{Madan}, {\it Let $K(x,y)$ be a function with real, complex or quaternion values, such that $K^*(x,y)=K(y,x),$ where $K^*=K$ if $K$ is real, $K^*$ is the complex conjugate of $K$ if it is complex, and  $K^*$ is dual of $K$ if it is a quaternion. Assume that
\beq
\int K(x,y)K(y,z) dy=K(x,z)+\lambda K(x,z)-K(x,z)\lambda,
\eeq
or symbolically
\beq
K*K=K+\lambda K-K\lambda
\eeq
with $\lambda$ a constant quaternion. Let $[K(x_i,x_j)]_N$ denote the $N\times N$ matrix with its $(i,j)$ element equal to $K(x_i,x_j).$ Then
\beq
\int \det[K(x_i,x_j)]_N dx_N=(A-N+1) \det[K(x_i,x_j)]_{N-1},
\eeq
 where $ A=\int K(x,x) dx$.  When $K(x,y)$ real or complex the variable $\lambda$ vanishes.}\\

 We further observe that the elements of the determinant in the \Eq{gt} is satisfied
\beqa
\int_0^{2\pi} \dfrac{\sin(N(t_j-t_j)/2)}{2\pi\sin((t_j-t_j)/2)}dt_j&=&N,\\
\int_0^{2\pi}\dfrac{\sin(N(t_j-t_l)/2)}{2\pi\sin((t_j-t_l)/2)}.\dfrac{\sin(N(t_l-t_ n)/2)}{2\pi\sin((t_l-t_n)/2)}dt_l&=&\dfrac{\sin(N(t_j-t_n)/2)}{2\pi\sin((t_j-t_n)/2)},
 \eeqa
and $A=N$ is a constant. Therefore, according to the theorem, the $M-$partical non-local correlation function can be written as
\beqa\label{non-local}
g_{M}(t_1,\ldots,t_M)
&=& \dfrac{(2\pi)^M}{L^M}\Big|\det\Big(\dfrac{\sin(N(t_j-t_l)/2)}{2\pi\sin((t_j-t_l)/2)}\Big)\Big|_{j,l=1,\ldots,M},\nn\\
g_{M}(x_1,\ldots,x_M) &=&\Big|\det\Big(\dfrac{\sin(N\pi(x_j-x_l)/L)}{L\sin(\pi(x_j-x_l)/L)}\Big)\Big|_{j,l=1,\ldots,M},
\eeqa
which can be used for calculating different nonlocal correlation functions of Tonks-Girardeau Bose gas.

\subsection{III.2 ~~Subleading contribution in $2$-body correlation $g_2$}

The Forrester et.al.~\cite{PJForrester_2006} have derived  the $O(1/c)$ oder of correction to the two-body non-local correlation function by expanding the momentum of bosons in $1D$ Lieb-Liniger model with a finitely strong repulsion.
This subleading contribution  can also be retrieved  by using the  fractional statistics following their analytical process.
When we consider the non-local correlations, the positions of particles (i.e. the permutation orders in the wave function) are very important.  In  the local correlation function, we used the  GES, $k_i=\alpha k_i^F$, the coordinates of the particles are demanded by a rescaling $x_i\sim x^F_i/\alpha$.
In contrast, such a rescaling of the particles' coordinates in a calculation of  the non-local correlation function works only for a strong repulsion limit, i.e. $x_i/c\ll 1$.
For calculating the  non-local correlation function, we  will use Forrester et.al.~\cite{PJForrester_2006} expansion method in term of the statistical parameter $\alpha$.

 For our convenience, we fix the  number of  total particles $N+2$ in the domain $0\leq y\leq x_1\leq \ldots\leq x_j\leq x\leq x_{j+1}\leq \ldots\leq x_N\leq L$ and label the particle positions as $(y,x_1,\ldots,x_j,x,x_{j+1},\ldots,x_N)=(\tilde{x}_1,\tilde{x}_2,\ldots,\tilde{x}_{N+2})$. Then, two-body non-local correlation function   reads
\beq
g_2^{N+2}(y,x)=\frac{(N+2)!}{N!}\dfrac{\int_{0}^L \ldots \int_0^L dx_1,\ldots,dx_N
|\psi_{N+2}|^2}{\int_0^L \ldots \int_0^L dy \,d x\, d{x}_1,\ldots,d{x}_{N}|\psi_{N+2}|^2}.
\eeq
For our convenience in notation,  we can denote the wave function in \Eq{GSE_x2} as
\beqa
\psi_{N+2}&\equiv &\psi(y,x_1,\ldots,x_j,x,x_{j+1},\ldots,x_N)=
 \psi(\tilde{x}_1^F,\ldots,\tilde{x}_{N+2}^F) \nonumber\\
 &=&
  \Big(1+(\alpha-1)\sum\limits_{l=1}^{N+2} \tilde{x}_l^F\dfrac{\partial}{\partial{\tilde{x}_l^F}}+\frac{1}{c}\sum\limits_{l=1}^{N+2}(2l-(N+2)-1)\dfrac{\partial}{\partial{\tilde{x}_l^F}}\Big)\psi_{N+2}^F.
\eeqa
Thus,  the operators in  the wave function can be expanded as
\beqa
\sum\limits_{l=1}^{N+2}\tilde{x}_l^F\dfrac{\partial}{\partial \tilde{x}_l^F}&=&y\dfrac{\partial}{\partial y}+x\dfrac{\partial}{\partial x}+\sum\limits_{l=1}^{N} x_l\dfrac{\partial}{\partial{x_l}},\\
\sum\limits_{l=1}^{N+2}(2l-(N+2)-1)\dfrac{\partial}{\partial \tilde{x}_l^F}&=&-(N+1)\dfrac{\partial}{\partial y}+\sum\limits_{l=1}^{j}(2l-N-1)\dfrac{\partial}{\partial x_l}+(2(j+1)-N-1)\dfrac{\partial}{\partial x}\nonumber\\
&&+\sum\limits_{l=j+1}^{N}(2(l+1)-N-1)\dfrac{\partial}{\partial x_l}.\label{decomposion}
\eeqa

Let $y=0$, and $\psi_{N+2}^F$ depends  only on $x_1,\ldots,x_N$ variables.
At the ground state, the system has zero total momentum, i.e.
 $\frac{\partial }{\partial y}=-\Big(\frac{\partial }{\partial x}+\sum_{l=1}^N\frac{\partial }{\partial x_l}\Big)$.
The wave function of free fermions  vanishes  when  two particles  coincide, i.e. $\psi_{N+2}^F$=0 at $x_l=0,L,x_j(j\neq l)$. Therefore,
\beqa
\int_{\tilde{x}_1\leq\ldots \leq \tilde{x}_j\ldots \leq\tilde{x}_{N+2}} \;dx_l\;\dfrac{\partial}{\partial x_l}|\psi_{N+2}^F|^2&=&0,\quad\text{and }\\
\int_{\tilde{x}_1\leq\ldots \leq \tilde{x}_j\ldots \leq\tilde{x}_{N+2}} \;dx_l \;x_l\dfrac{\partial}{\partial {x}_l}|\psi_{N+2}^F|^2&=&-\int_{\tilde{x}_1\leq\ldots \leq \tilde{x}_j\ldots\leq\tilde{x}_{N+2}} |\psi_{N+2}^F|^2 dx_l.
\eeqa
Taking  together  these conditions with the \Eq{decomposion}, we may show  that $\mathcal{N}^2=\alpha^{-(N+1)}(N+1)!L^{N+1}$
and
\beq\label{g_n2}
g_2^{N+2}(0,x)
=\alpha^{(N+1)} \Bigg[1+(\alpha-1)\Big(x\dfrac{\partial}{\partial x}-N \Big)\Bigg]g_2^{N+2,F}(0,x)+\frac{1}{\mathcal{N}^2}\dfrac{2}{c}\int\ldots\int_{\tilde{x}_1\leq\ldots\leq\tilde{x}_{N+2}} dx_1,\ldots,dx_N(j+1)\dfrac{\partial}{\partial x}\psi_{N+2}^F(0,x),
\eeq
where $\psi_{N+2}^F(0, x_1,\ldots,x,\ldots,x_N)\equiv \psi_{N+2}^F(0,x)$.

Furthermore, the coefficient of $g_2^{N+2,F}(0,x)$ in \Eq{g_n2} can be simplified in term of the GES parameter $\alpha$.
 \Eq{g_n2} has been derived in terms of $1/c $ corrections  ~\cite{PJForrester_2006}. A detailed calculation of the integral term (in  {\bf Appendix C}), gives
\beq
\frac{1}{\mathcal{N}^2}\int\ldots\int_{\tilde{x}_1\leq\ldots\leq\tilde{x}_{N+2}} (j+1) dx_1,\ldots,dx_N\dfrac{\partial}{\partial x}\psi_{N+2}^F(0,x)=\alpha^{N+1}\Big(\dfrac{\partial}{\partial x}g_2^{N+2,F}(0,x)+
\dfrac{\partial}{\partial x}\int_0^{x} dx_l\;g_3^{N+2,F}(0,x,x_l) \Big),
\eeq
where $g_2^{N+2,F}(0,x)$  and $g_3^{N+2,F}(0,x,x_l)$ are  the two-body and three-body non-local correlation functions of $N+1$ free fermions,  which have been  given in \Eq{non-local}. Finally, the \Eq{g_n2} can be viewed as the two-body  non-local correlation function of $N$ particles, namely
\beq\label{g2_non_local}
g_2^{N}(0,x)=\alpha g_2^{N,F}(0,x)+(\alpha-1)x\dfrac{\partial}{\partial x} g_2^{N,F}(0,x)+\dfrac{2}{c}\dfrac{\partial}{\partial x}g_2^{N,F}(0,x)+
\dfrac{2}{c }\dfrac{\partial}{\partial x}\int_0^{x} dx_l\;g_3^{N,F}(0,x,x_l).
\eeq
We observe that the two-body correlation function essentially depends on  the  two-body and higher order  non-local correlation functions of  the free fermions and their derivatives, which  provide insights into the many-body correlations in strongly interacting bosons.

\section{ IV. Large momentum Tail of the momentum distribution and Tan's Contact relations}

Various methods can be used to calculate  the contact \cite{Tan-coantact,Tan1:2008,Tan2:2008,zhang:2009,Werner:2009,Braaten:2019}, for example, it can be obtained from the free energy or ground state energy  via the Helmann-Feynmann theorem~\cite{Ovidiu_2017}, or from the single particle correlation function~\cite{Joaquin}, from the operator product expansion field theory method~\cite{Barth:2011}, etc.

\subsection{IV.1 ~Large momentum Tail of the momentum distribution}

The single-particle momentum distribution, $w(p)$ at momentum $p=\hbar k$,  is the Fourier transform of the one-particle density matrix $\rho(x_j,x_k)$.
In order to build up a connection between the Tan's contact and the power law decay with distance in large momentum tail of  momentum distribution, we first calculate the wave function in momentum space.
To this end, we introduce the centre of mass coordinate, $R=(x_j+x_k)/2$ and relative coordinate, $x_{kj}=x_k-x_j,$ of the $j,k$ pair of  particles with other fixed $N-2$ coordinates $\{x_i\}_{i\neq j,k}$.
The $k^{th}$ particle in the neighbor of $j^{th}$ particle, such as $x_{1}\leq x_{2}\leq x_{3}\leq\ldots\leq x_{k-1}\leq x_{j}\leq x_{k}\leq x_{k+1}\leq\ldots\leq x_{N}$  and  $x_{1}\leq x_{2}\leq x_{3}\leq\ldots\leq x_{k-1}\leq x_{k}\leq x_{j}\leq x_{k+1}\leq\ldots\leq x_{N}$, the wave function for 1D strongly interacting bosons in ~\Eq{wave} can be approximately written as
\beq\label{wave2}
\psi^B(x_1,\ldots,x_N)=\dfrac{1}{\mathcal{N}} \mathcal{A}(c)\sum\limits_p(-1)^p e^{\I(k_{p_{k}}+k_{p_{j}})R}\Big(1+\frac{\I}{2}(k_{p_{k}}-k_{p_{j}})x_{kj}-\frac{1}{8}(k_{p_{k}}-k_{p_{j}})^2x_{kj}^{2}+O(x_{kj}^3)\Big)e^{\I\sum\limits_{i\neq k,j;i=1}^{N}k_{p_{i}}x_i}.
\eeq

Following symmetric conditions, the \Eq{wave2} can be represented in general forms as
\beqa\label{B-wave}
\psi^B(x_1,\ldots,x_N)\Big|_{x_{k}=x_{j}^{+}} &\approx &\dfrac{1}{\mathcal{N}}\Big(1+\frac{1}{2}\Big(\frac{\partial}{\partial x_{k}}-\frac{\partial}{\partial x_{j}}\Big)|x_{kj}| sgn(x_{kj})+\frac{1}{8}\Big(\frac{\partial}{\partial x_{k}}-\frac{\partial}{\partial x_{j}}\Big)^{2}x_{kj}^2\Big)\psi,
\eeqa
for bosons in the $1D$ Lieb-Liniger model,
where $sgn(x)$ is $+1$ for $x>0$ and $-1$ for $x<0$,
$\psi$ that corresponds to the wave function  \Eq{wave} at ${x_{k}=x_{j}^{+}}$.

The Fourier transform of the one-particle correlation function is the momentum distribution:
\beq
w(p)=\int dx\;dx^\prime g_1(x;x^\prime)e^{-\I p(x-x^\prime)},
\eeq
where $p=(2\pi/L)s$ and $s$ is an integer.
This is equivalent to calculating first the many-particle wave function in momentum space for one atom and then integrating out the rest of the coordinates from its square modulus.
In order to calculate the asymptotic behaviour (the large $p$ tail) of the momentum distribution \cite{Rafael},
we  first consider the normalized wave function with respect to the first particle in momentum space
\beq\label{psip}
\psi^B(p,x_{2},\ldots,x_k,\ldots,x_{N})=\sum\limits_{k=2}^N\frac{1}{\sqrt{L}}\int_{0}^{L}dx_{1}\;e^{-ipx_{1}/\hbar}\psi^B(x_{1},x_{2},\ldots,x_k,\ldots,x_{N}),
\eeq
where $p=(2\pi\hbar/L)l$ where $l$ is an integer.
For a repulsive interaction, the diagonal terms becomes much larger than the off-diagonal terms  due to the periodic boundary condition and symmetry.
 For a periodic function $F_{1}(x_{1},\ldots,x_{N})\;(x_k-x_1)\; sgn(x_k-x_1)$, defined on the interval $[0,L]$, where
$F_{1}(x_{1},\ldots,x_{N}) =\big(\frac{\partial}{\partial x_{k}}-\frac{\partial}{\partial x_{1}}\big)\sum\limits_P(-1)^P exp(\I\sum_{j=1}^{N}k_{P_{j}}x_j)$  is a regular function, we thus rewrite the  square modulus as
\beqa
|\psi^B(p,x_{2},\ldots,x_k,\ldots,x_{N})|^2
&\approx&\frac{(N-1)}{\mathcal{N}^2L(p/\hbar)^4}\Big(1+\frac{1}{c}\sum\limits_{j=1}^{N}(2j-N-1)\frac{\partial}{\partial x_j}\Big)\Big|F_{1}(x_{1},\ldots,x_{N})\Big|^{2}\Bigg|_{x_k=x_1}+O\Big(\frac{\hbar^5}{p^5}\Big).
\eeqa
After a length calculation, the  asymptotic behaviour of the momentum distribution is given by
\beqa\label{wp}
w(p)&\overset{p\to\infty}{=}&N\int_{0}^{L}\ldots\int_0^L dx_2\ldots dx_k\ldots dx_N\big| \psi^B(p,x_{2},\ldots,x_k,\ldots,x_{N})\big|^{2}\label{WP}
\\
&=&N\int_{0}^{L}\ldots\int_0^L dx_2\ldots dx_N\;\Bigg[\frac{(N-1)}{(p/\hbar)^4\mathcal{N}^2L}\Big(1+\frac{1}{c}\sum\limits_{j=1}^{N}(2j-N-1)\frac{\partial}{\partial x_j}\Big)\Big| F_{1}(x_{1},\ldots,x_{N})\Big|^{2}\Bigg]\Bigg|_{x_k=x_1^{+}}\nn\\
&=&
\frac{N(N-1)}{(p/\hbar)^4\mathcal{N}^2L}\int_{0}^{L}\ldots\int_0^L dx_2\ldots dx_N\Bigg[\Big( 1
+\frac{1}{c}\sum\limits_{j=1}^{N}(2j-N-1)\frac{\partial}{\partial x_j}\Big)\Big|F_{1}(x_{1},\ldots,x_{N})\Big|^{2}+O\Big(\frac{1}{c^2}\Big)
\Bigg]\Bigg|_{x_k=x_1^{+}}.
\eeqa


Applying the GSE relation, $k_{P_{j}}=k_{P_{j}}^{F}\alpha$, making a rescaling $x_j=x_j^{F}/\alpha$, and after a lengthy calculation, we obtain the momentum distribution of $1D$ bosons up to order $1/c$
\beqa
w(p)&\overset{p\to\infty}{\approx}&
\frac{\alpha^{2}}{L(p/\hbar)^4}\int_{0}^{\alpha L}dx_k^{F}\Big(1+\frac{\alpha}{c}\sum\limits_{j=1}^{N}(2j-N-1)\frac{\partial}{\partial x_j^F}\Big)\Big(\frac{\partial}{\partial x_{k}^{F}}-\frac{\partial}{\partial x_{1}^{F}}\Big)\Big(\frac{\partial}{\partial y_{k}^{F}}-\frac{\partial}{\partial y_{1}^{F}}\Big)g_2^{ F}(x_1^F,x_k^F;y_1^F,y_k^F)\Big|_{x_k^F\to x_1^{F+}},
\label{wp1}
\eeqa
where $g_2^F$ is fermions two-particle correlation, 
Up to order $1/c$, the normalization factor for strongly interacting $1D$ bosons has been calculated in~\cite{HCF} as
$
\mathcal{N}^2=\alpha^{1-N}\int_0^{ L}\ldots\int_0^L dx_1^F\ldots dx_N^F\big|\sum\limits_P(-1)^P exp\big(\I\sum\limits_{j=1}^{N}k^F_{P_{j}}x^F_j\big)\big|^2\Big|_{x^F_k=x^F_1}.
$

By applying Wicks theorem and getting partial derivatives, the asymptotic behaviuor of the momentum distribution becomes,
 \beq\label{tail}
 w(p)\overset{p\to\infty}{=} \frac{g_2(0,0)\gamma^2 n^2}{(p/\hbar)^4}=\frac{4\alpha^{3}\pi^2n^4}{3(p/\hbar)^4}+O\Big(\frac{1}{c^2},\frac{1}{(p/\hbar)^5}\Big).
 \eeq
 In the above  calculation \Eq{wp1}, we have proved that the leading order part  gives the universal asymptotic behaviour \Eq{tail}, whereas the second term in \Eq{wp1} is zero.
 According to the definition of momentum distribution, the \Eq{tail} is equal to the relation  given in ~\cite{Maxim_2003} and ~\cite{Maxim1_2003}.
Moreover, the Hellmannn-Feynman theorem relates to the derivative of the energy with respect to the interacting strength.
So  we have a  two-particle local correlation function $g_2^B(x_1,x_1)=n^2 e^{\prime}(\gamma)$
\beq
w(p)\label{result2}
\overset{p\to\infty}{=}\dfrac{ e^{\prime}(\gamma)n^4\gamma^2}{ (p/\hbar)^4},
\eeq
where the derivative of total energy, $e^{\prime}(\gamma)\approx\frac{\hbar^2}{2m}\frac{\pi^2\alpha^2}{3}$~\cite{Guan_2007}.
The result of  \Eq{tail}  shows  that the tail of the momentum distribution decays with a power of $1/(p/\hbar)^4$  and it has a sub-leading contribution determined by the GES parameter $\alpha$.

\section{IV.2 ~The contact relation}

A powerful universal relation  connects the strength of short-range two-particle correlations to the thermodynamics of a many-particle system with zero-range interaction~\cite{Wild_2012}.
The  contact allows for a simple analytic expression that unveils how to scale the momentum distribution function at large momentum~\cite{Xu:2015}.
We recall  the {\bf Lemma} discussed  in \cite{Maxim_2003,Ovidiu_2017} and ~\cite{Patu:2017}.\\

{\bf Lemma 1} {\it Consider a function absolutely integrable, vanishing at infinity which has a singularity of the type $f(x)=|x-x_0|^\alpha F(x)$ at $x_0$ with $F(x)$ analytical and $\alpha>-1,\alpha\neq,0,2,4,\ldots$. Then we have

\beq\label{lemma}
\lim\limits_{k\to\infty}\int_{-\infty}^{\infty}e^{-\I kx}f(x)dx\approx 2\cos\frac{\pi}{2}(\alpha+1)\tau(\alpha+1)\dfrac{e^{-\I kx_0}}{|k|^{\alpha+1}}F(x_0)+O(1/|k|^{\alpha+2}).
\eeq

In the case of multiple singular points of the same type, the asymptotic behaviour of the integral is given by the sum of all the corresponding contributions given by the right hand side of the \Eq{lemma}}.\\

Considering discontinuity of the  derivative wave function, the \Eq{B-wave}  is given by
\beqa\label{B-wave1}
\psi^B(x_1,\ldots,x_N)
&=&\dfrac{1}{\mathcal{N}}\Big(1-\dfrac{|x_{kj}|}{a_{1D}}sgn(x_{kj})+\dfrac{2x_{kj}^2}{a_{1D}^2}+O(x_{kj})^3\Big)\psi,
\eeqa
where $c=-2/a_{1D}$.
Applying the Lemma 1 to the \Eq{B-wave1} and after changing integration variables  as $dx_1\sim dx$ and $dx_2\sim dR$ at $x_1\to x_k$, we find that the function $\psi$ is a function of $R$ and $X$, where the set of fixed coordinates $X=\{x_i\}_{i=2\ldots,k-1,k+1,\ldots,N}$.  Thus we may calculate  the expression  \Eq{psip} as
\beqa
|\psi^B(p,x_{2},\ldots,x_k,\ldots,x_{N})|^2&\approx &
\dfrac{(N-1)}{\mathcal{N}^2L} \Big[\int_0^L dx e^{-\I px/\hbar}\Big(1-\dfrac{|x|}{a_{1D}}\Big)|\psi(R,R,X)|\Big]^2\nonumber\\
&\overset{{p\to\infty}}{=}&
\dfrac{(N-1)}{\mathcal{N}^2L}\Big(\dfrac{4}{a_{1D}^2( p/\hbar)^4}\Big)|\psi(R,R,X)|^2.
\eeqa
In the above equation,
we have neglected the off-diagonal terms which vanish faster than the diagonal terms.
 From \Eq{WP}, the large momentum  tail of the momentum distribution  is obtained as
\beqa
w(p)&\overset{{p\to\infty}}{=}&
\dfrac{4}{a_{1D}^2L( p/\hbar)^4}C_2,
\eeqa
where  the two-particle contact defined by $C_2=\int_0^L dR\; g_2(R,R)$.
Here the  two-particle contact  is given by $C_2=4\pi^2nN\gamma^{-2}\alpha^3/3$.
Whereas  the two-particle local correlation function reads   $g_2(0,0)=4\pi^2n^2\alpha^3/(3\gamma^2)$ for a finitely strong repulsive Bose gas  at zero temperature, further confirming  \Eq{s2body}.
This result agrees with  the result which has been derived by Yuta et.al. in~\cite{Yuta}.
Moreover, in a weakly coupling regime, the two-particle local correlation functions is given by  $g_2(0,0)=n^2(1-2\sqrt{\gamma}/\pi)$.
 So the two-particle contact can be obtained as $C_2=nN(1-2\sqrt{\gamma}/\pi)$ for the 1D weakly interacting bosons.

\section{IV.3~ Static Structure factor}

The structure factor is a property that defines how an ensemble of atoms scatters incident radiation.
 Experimentally, it is usually measured by two-photon Bragg scattering.~\cite{Rafael}.
Due to the translational invariance of the system, the density is constant ~\cite{Yuta}, the static structure factor is  defined as
\beq
S(k)=1+\dfrac{1}{N}\int dx_k\int dx_j\; e^{-\I k(x_k-x_j)}\Big(g_2(x_k,x_j)-n(x_k)n(x_j)\Big),
\eeq
where $n(x_i)$ is the density.  Changing integration variables  as before, we have
\beqa
S(k)
&=&1+\dfrac{1}{N}\int_0^L dx \int_0^L dR\; e^{-\I kx}\left(\left( 1-\dfrac{2|x|}{a_{1D}}\right)g_2(R,R^+)-n^2\right)\label{sfactor}\\
&=&1+\dfrac{4}{Na_{1D}k^2}C_2,\label{s1}
\eeqa
where $C_2=\int_0^L dR\; g_2(R,R^+)$, also see  ~\cite{Yuta}.
By using \Eq{non-local}, the two-particle non-local correlation function of the 1D Bose gas in strongly interacting  regime, i.e. $c\to\infty$,  is given by
\beqa
g_{2}(R,R^+) &\approx &n^2\Big[1-\Big(\dfrac{\sin(k^F x)}{k^F x}\Big)^2\Big].
\eeqa
Here, $x$ and $k^F=\pi n$ are the  relative coordinate and  the  cut-off Fermi  momentum.
Thus  the structure factor can be determined by  \Eq{sfactor}, namely
\beq
S(k)=
\begin{cases}
1;&\mbox{ for}\;|k|>2k^F\\
\dfrac{|k|}{2k^F};&\mbox{ for}\;|k|<2k^F.
\end{cases}
\eeq

In fact,  in the low-momentum and the infinitely repulsive regime,  the static structure factor  has  a linear dependence of  $k$.
In the TG regime this behaviour converts to $S(k)=|k|/2k^F$.
For high momentum, the structure factor always converges to $1$.
%


\section{V. ~Conclusion}
In summary, we have presented a pedagogical study of various higher order local and non-local correlation functions of 1D  Lieb-Liniger Bose gas.
Using the Bethe wave function, we have studied  the non-local $2M$ point correlation functions and large momentum tail of momentum distributions  in term of the fractional statistical parameter $\alpha$ for the system with a strong repulsion.
We have also discussed the higher order correlation functions for the weakly interacting Bose gas via the Bethe ansatz wave function.
It turns out  that the  leading order of the large-momentum tail is  determined by the contact, which is related to the short-distance behaviour of the two-body density matrix.
The  two-body density matrix  has been given in term of the fractional statistical parameter $\alpha$, giving  the  subleading order $1/c$ correction in the strong coupling regime, where the interaction strength $c \gg1$.
Our results show that the fractional statistical parameter $\alpha$ attributes  to the sub-leading contributions of the local, non-local correlation functions, two-body  contact and static structure factor, etc.
Our method is applicable to other exactly solvable models.

\paragraph{\bf Acknowledgments:}
We thank Xiaoguo Yin, Yu-Zhu Jiang, Shina Tan for helpful discussion. 
This work is supported by the NSFC under grant numbers 11374331,  the key NSFC grant No.\ 11534014 and the National Key R\&D Program of China  No. 2017YFA0304500.
This work  has been partially supported   by CAS-TWAS President's Fellowship for International PhD students.

\section{Appendix A}
The Betha ansatz wave function  given in \Eq{BA_wave}
\beq
 \psi(x_1,\ldots,x_N)=\sum\limits_P(-1)^P\Big[\prod\limits_{1\leq i< j\leq N}\Big(\dfrac{\I c+k_{P_i}-k_{P_j}}{\I c-k_{P_i}+k_{P_j}}\Big)^{1/2}\Big] e^{\I \sum\limits_{j=1}^Nk_{P_{j}}x_j}.
 \eeq
Using  the Bethe ansatz equation, we have
\beqa
(-1)^PA(P)\equiv(-1)^P\prod_{i<j}\Big(\dfrac{\I c+k_{P_i}-k_{P_j}}{\I c-k_{P_i}+k_{P_j}}\Big)^{1/2}
&=&(-1)^P\prod_{i<j}\Big[-2Tan^{-1}\Big(\dfrac{k_{p_i}-k_{p_j}}{c}\Big)\Big]^{1/2}\nn\\
&\equiv&(-1)^P\prod_{i<j}\Big(e^{\I\theta(k_{p_i}-k_{p_j})}\Big)^{1/2}\nn\\
&=&e^{\I\frac{\pi}{2}(1-(-1)^P)}\;e^{\frac{\I}{2}\sum_{i<j}\theta(k_{p_i}-k_{p_j})}\nn\\
&=&exp\Big(\I\frac{\pi}{2}(1-(-1)^P)+\frac{\I}{2}\sum_{i<j}\theta(k_{p_i}-k_{p_j})\Big).\label{weak_ap}
\eeqa
Furthermore we consider  the following  relation
\beqa\label{weak_theta}
Arctan(x)+Arctan(1/x)=\begin{cases}
\pi/2\quad \text{for}\;x>0,\\
-\pi/2\quad \text{for}\;x<0.
\end{cases}
\eeqa

Then, we have
\beqa
\theta(k_{i}-k_{j})=-2Tan^{-1}\Big(\dfrac{k_i-k_j}{c}\Big)
&=&-\dfrac{(k_i-k_j)}{|k_i-k_j|}\pi+\dfrac{2c}{k_i-k_j}.
\eeqa
Taking $q_j=\sqrt{\dfrac{L}{2c}}k_j$,
\beqa
\theta(k_{i}-k_{j})=\theta(\sqrt{\dfrac{2c}{L}}(q_i-q_j))
&=&-\dfrac{(q_i-q_j)}{|q_i-q_j|}\pi+\sqrt{2cL}\;F(q_i-q_j),
\eeqa
where $F(q_i-q_j)=(q_i-q_j)^{-1}$. Therefore, \Eq{weak_ap} becomes
\beq\label{pap}
(-1)^PA(P)=exp\Big(\I\frac{\pi}{2}(1-(-1)^P)+\frac{\I}{2}\sum_{i<j}\Big[-\dfrac{(q_i-q_j)}{|q_i-q_j|}\pi+\sqrt{2cL}\;F(q_i-q_j)\Big]\Big).
\eeq
 Using \Eq{pap}, we consider the two permutations of the system with $3$ particles.
\beqa
(-1)^PA(123)&=&exp\Big(\I\frac{\pi}{2}(1-1)+\frac{\I}{2}\sum_{i<j}\Big[-\dfrac{(q_i-q_j)}{|q_i-q_j|}\pi+\sqrt{2cL}\;F(q_i-q_j)\Big]\Big)\nn\\
&=&exp\Big(\frac{\I}{2}3\pi\Big)exp\Big(\I\sqrt{\dfrac{cL}{2}}\Big[F(q_1-q_2)+F(q_1-q_3)+F(q_2-q_3)\Big]\Big).
\eeqa

\beqa
(-1)^PA(213)&=&exp\Big(\I\frac{\pi}{2}(1-(-1))+\frac{\I}{2}\sum_{i<j}\Big[-\dfrac{(q_i-q_j)}{|q_i-q_j|}\pi+\sqrt{2cL}\;F(q_i-q_j)\Big]\Big)\nn\\
&=&exp\Big(\frac{\I}{2}3\pi\Big)exp\Big(\I\sqrt{\dfrac{cL}{2}}\Big[F(q_2-q_1)+F(q_2-q_3)+F(q_1-q_3)\Big]\Big).
\eeqa
Therefore, the general form of  amplitude with permutation's sign of the wave function in the weak coupling regime can be obtained as
\beq\label{general_amplitude_weak}
(-1)^PA(p)=exp\Big(\I\sqrt{\dfrac{cL}{2}}\sum_{i<j}\Big[\;F(q_i-q_j)\Big]\Big).
\eeq
Finally,  we rewrite the Bethe ansatz wave function, \Eq{wave}  in the weak coupling regime as
\beqa
\psi(x_1,\ldots,x_N)
&=&\sum\limits_P\;exp\left(\I\sqrt{\dfrac{cL}{2}}\sum_{i<j}\left[\sqrt{2cL}\;F(q_{P_i}-q_{P_j})\right]\right)exp\Big(\I\sum\limits_{j=1}^N\sqrt{\dfrac{2c}{L}}q_{P_j}x_j\Big)\nn\\
&=&\sum\limits_P\Bigg[1+\I\Big(\sqrt{\dfrac{cL}{2}}\sum_{i<j}F(q_{P_i}-q_{P_j})+\sqrt{\dfrac{2c}{L}}\sum\limits_{j=1}^Nq_{P_j}x_j\Big)\nn\\
&&\quad-\dfrac{1}{2}\Big(\sqrt{\dfrac{cL}{2}}\sum_{i<j}F(q_{P_i}-q_{P_j})+\sqrt{\dfrac{2c}{L}}\sum\limits_{j=1}^Nq_{P_j}x_j\Big)^2+\ldots\Bigg]\nn\\
&=&N!-\dfrac{c}{2}Y(x_1,\ldots,x_N),\label{W-wave}
\eeqa
 where
  $Y(x_1,\ldots,x_N)=\sum\limits_P\Big(\sqrt{\dfrac{L}{2}}\sum_{i<j}F(q_{P_i}-q_{P_j})+\sqrt{\dfrac{2}{L}}\sum\limits_{j=1}^Nq_{P_j}x_j\Big)^2
$,
since the symmetric condition gives
$\sum\limits_P \sum_{i<j}\;F(q_{P_i}-q_{P_j})=0$ and $\sum\limits_P\sum\limits_{j=1}^Nq_{P_j}=0$.\\

\section{Appendix B}

Let us represent the field operator $\hat\Psi$ as a sum of the condensate part $\psi_{0}$ and a non-condensate part $\psi^{\prime}$ such as
 \beq\label{p0_p1}
 \hat\Psi =\psi_{0}+\psi^{\prime}.
 \eeq
In the uniform case, we may let  the condensate contain a macroscopic number of particles $N_0$.
 Consider the density of the system is
$ n=n_0+n^\prime$
where $n_0=\langle\psi_0\psi_0^*\rangle=|\psi_0|^2$ is the condensed particles density and $n^\prime=\langle\psi^{\prime\dagger}\psi^\prime\rangle$ is the non condensed particles density.
Considering the $M^{th}$ order density of the system  with respect to the first order of  $\psi^\prime$, we have
\begin{align}
 n^M&=(|\psi_0|^2)^M+M(|\psi_0|^2)^{M-1}\langle\psi^{\prime\dagger}\psi^\prime\rangle+O\big(\langle\psi^{\prime\dagger}\psi^\prime\rangle\big)^2,\\
\Rightarrow\; (|\psi_0|^2)^M&=n^M -M(|\psi_0|^2)^{M-1}\langle\psi^{\prime\dagger}\psi^{\prime}\rangle\nn\\
&= n^M-Mn^{M-1}\langle\psi^{\prime\dagger}\psi^{\prime}\rangle+O\big(\langle\psi^{\prime\dagger}\psi^\prime\rangle\big)^2.\label{eqp}
 \end{align}

 The $M$ particle correlation function for $1D$ bosons at weak coupling limit reads
 \beq\label{gm_weak}
 g_M^w=\big<(\psi_0^{\ast}+\psi^{\prime\dagger})^M(\psi_0+\psi^{\prime})^M\big>.
 \eeq
  Expanding the $M^{th}$ order in \Eq{gm_weak} and substituting \Eq{eqp} into the above equation, we obtain
 \begin{align}
 g_M^w
 &=\big<\big(\psi_0^{\ast M}+M\psi_0^{\ast(M-1)}\psi^{\prime\dagger}+ \frac{1}{2}M(M-1)\psi_0^{\ast(M-2)}\psi^{\prime\dagger2} +O(\psi^{\prime\dagger 3})\big)\nn\\
 &\hspace{3cm}\times
 \big(\psi_0^M+M\psi_0^{M-1}\psi^{\prime} + \frac{1}{2}M(M-1)\psi_0^{(M-2)}\psi^{\prime2}
 +O(\psi^{\prime 3})\big)\big>\nn\\
&=\big<| \psi_0|^{2M}+M^2 | \psi_0|^{2(M-1)}\big(\psi^{\prime\dagger}\psi^\prime\big)+M(M-1) | \psi_0|^{2(M-1)}\psi^{\prime2}\big>\nn\\
  &= n^M\Big(1+\frac{M(M-1)}{n}\big(\langle\psi^{\prime\dagger}\psi^{\prime}\rangle+\langle\psi^\prime\psi^\prime\rangle\big)\Big).\label{eq}
 \end{align}
 \Eq{eq} shows that $g_M^w$  depends on the normal density of non-condensed particles,$\langle\psi^{\prime\dagger}\psi^{\prime}\rangle$, and anomalous density of non-condensed particles $\langle\psi^\prime\psi^{\prime}\rangle$.
Consider the form of  non-condensate part,$\psi^\prime,$ as
 \beq\label{bogoliubov}
 \psi^\prime=\sum\limits_\nu\Big(u_\nu\hat b_\nu e^{-\I\epsilon_\nu t/\hbar}-v_\nu^{\ast}\hat b_\nu^\dagger e^{\I\epsilon_\nu t/\hbar}\Big),
 \eeq
which  is known as  Bogoliubov transformation.  Here  the index $\nu$ labels quantum states of elementary excitations, $\hat{b}_\nu,\hat{b}_\nu^\dagger$ are operators of excitations, $\epsilon_\nu$ are their eigenenergies, and $u_\nu,v_\nu$ their eigen functions.
 The operators $\hat b_\nu, \hat b_\nu^\dagger$ obey the usual boson commutation relations 
 \beq
\hat b_\nu \hat b_{\nu^\prime}^\dagger-\hat b_{\nu^\prime}^\dagger\hat b_\nu =\delta_{\nu\nu^\prime},\quad \quad
\hat b_\nu \hat b_{\nu^\prime}-\hat b_{\nu^\prime}\hat b_\nu =0.
 \eeq
The functions $u_\nu,v_\nu$ are normalized by the condition~\cite{dmg_gvs}
 \beq\label{weak_normalization}
 \int (u_\nu u_\nu^*-v_\nu v_\nu^*) dx =\delta_{\nu\nu^\prime}.
 \eeq

Taking into account that the index $\nu$ is now the excitation wave vector $k$, we write the excitation wave  function in the form
 $
 u_\nu=\frac{u_k}{\sqrt{L}}e^{\I kx}\quad \text{and}\quad v_\nu=\frac{v_k}{\sqrt{L}}e^{\I kx}.
 $
 Consequently, we have
 \begin{align}
  \psi^\prime
  &=\frac{1}{\sqrt{L}}\sum\limits_k\Big(u_k \hat b_k e^{\I(kx-\epsilon_k t/\hbar)}-v_k^{\ast}\hat b_k^\dagger e^{-\I(kx-\epsilon_k t/\hbar)}\Big). \label{p1}
 \end{align}

  The normal and anomalous density of the non-condensate part can be calculated by using \Eq{p1} at $k=k^\prime$ as
\begin{align}
\left\langle\psi^{\prime\dagger}\psi^{\prime}\right\rangle
&=\frac{1}{L}\sum\limits_{k}\Big[u_k^2 N_k+ v_{k}^2(1+N_k)\Big],\label{eq3}\\
\langle\psi^\prime\psi^{\prime}\rangle
&= \frac{1}{L}\sum\limits_{k}\Big[ -u_kv_k^\ast(1+N_k)-v_k^\ast u_k N_k\Big],\label{eq4}
 \end{align}
where $N_k$ is the occupation number for the excitations with
 $N_k\rightarrow 0$ as $T\rightarrow 0$ and commutation relations of the excitation operators  are
$\langle \hat b_{k^\prime}^
\dagger\hat b_k\rangle =\delta_{k^\prime k}N_k,\quad
\langle\hat b_{k^\prime}\hat b_k^\dagger \rangle=\delta_{k^\prime k}(1+N_k),\quad
\langle\hat b_{k^\prime} \hat b_k\rangle =\langle\hat b_{k^\prime}^\dagger\hat b_k^\dagger\rangle=0$.
 Summation of \Eq{eq3} and \Eq{eq4} ,
 \beq\label{sum_p0_p1}
 \left\langle\psi^{\prime\dagger}\psi^{\prime}\right\rangle+\langle\psi^\prime\psi^{\prime}\rangle=\frac{1}{L}\sum\limits_{k}\Big[u_k^2 N_k+ v_{k}^2(1+N_k)-u_kv_k(1+2N_k)\Big],
 \eeq
 since $v_k^\ast=v_k$.
 By considering up to the linear order of $\psi^\prime$ of field operator \Eq{p0_p1}, the Gross-Pitaevskii equation can be obtain as
 \beq\label{A_GP_eq}
 \I\hbar\dfrac{\partial\psi^\prime}{\partial t}=\Big(-\dfrac{\hbar^2}{2m}\dfrac{\partial^2}{\partial x^2}+2g|\psi_0|^2-\mu\Big)\psi^\prime+g\psi_0^2\psi^{\prime\dagger}.
 \eeq
 The non-condensate density yields from the eigenfunctions $u_k$ and $v_k$.
 In order to find $u_k$ and $v_k$, evaluate the commutator of both sides of \Eq{A_GP_eq} with $\hat{b}_k$ and repeating procedure with $\hat{b}_k^\dagger$, one can get

 \begin{align}
-  \epsilon_k  v_k&=\Big(\dfrac{\hbar^{2}k^2}{2m} +2g|\psi_0|^{2}-\mu\Big) v_k -g\psi_{0}^{2} u_{k}, \label{feq4}\\
\epsilon_k u_k &=\Big(\dfrac{\hbar^{2}}{2m}k^2+2g|\psi_0|^{2}-\mu\Big) u_k-g\psi_{0}^{2} v_{k}. \label{ffeq4}
 \end{align}
Here, $v^\ast=v\; \text{and} \; u^\ast=u$.
Substituting $\mu=n_0g,$ and
$E_k=\dfrac{\hbar^{2}k^2}{2m},$
to the \Eq{feq4} and \Eq{ffeq4} which are called  Bogoliubov-de Gennes equations for elementary excitations, are transformed to
    \begin{align}
  -  \epsilon_k  v_k  &=\big(E_k +gn_0\big) v_k -gn_0 u_{k},\label{feq05}\\
   \epsilon_k u_k& =\big(E_k+gn_0\big) u_k-gn_0 v_{k}.\label{feq06}
    \end{align}
  After taking difference between squares of \Eq{feq05} and   \Eq{feq06},  applying  the normalization condition given \Eq{weak_normalization}, $|u_k|^2 -|v_k|^2=1,$ then we obtain the excitation energy 
  \beq
  \epsilon_k=\sqrt{E_k^2 +2gn_0E_k}.\label{e1}
  \eeq
  Taking a summation of  \Eq{feq05} and   \Eq{feq06} gives
  \beq\label{uk}
  u_k=\dfrac{(E_k+\epsilon_k)v_k}{\epsilon_k-E_k}.
  \eeq
  Applying the  $|u_k|^2 -|v_k|^2=1$, the \Eq{uk} gives
  \beq\label{vk}
  v_k^2=\dfrac{(\epsilon_k-E_k)^2}{4E_k\epsilon_k}.
  \eeq
  From \Eq{uk} and \Eq{vk},  we finally obtain
  \bea
v_k=\pm\dfrac{1}{2}\Big(\sqrt{\frac{\epsilon_k}{E_k}}-\sqrt{\frac{E_k}{\epsilon_k}}\Big),
\quad\text{and}\quad
  u_k=\pm\dfrac{1}{2}\Big(\sqrt{\frac{\epsilon_k}{E_k}}+\sqrt{\frac{E_k}{\epsilon_k}}\Big).\label{eq7}
  \eea

Substituting  \Eq{eq7} into the   \Eq{sum_p0_p1}, and solving the summation of normal and anomalous density, then we have
  \beq\label{fluctuation}
  \langle\psi^{\prime\dagger}\psi^{\prime}\rangle+ \langle\psi^\prime\psi^{\prime}\rangle=\frac{1}{L}\sum\limits_{k}\Big[\frac{E_k}{\epsilon_k}(N_k+\frac{1}{2})-\frac{1}{2}\Big].
  \eeq

 The \Eq{total}  consists quantum vacuum fluctuations and thermal fluctuations of the system. At zero temperature all particles are in the condensate, i.e. $N_k=0$ at $t=0$ and $N_k\neq 0$ for finite temperature.  Therefore, density of non-condensate can be represented as
  \beqa\label{total}
     \langle\psi^{\prime\dagger}\psi^{\prime}\rangle+ \langle\psi^\prime\psi^{\prime}\rangle\Big|_{total}&=&\langle\psi^{\prime\dagger}\psi^{\prime}\rangle+ \langle\psi^\prime\psi^{\prime}\rangle\Big|_{T=0}+\langle\psi^{\prime\dagger}\psi^{\prime}\rangle+ \langle\psi^\prime\psi^{\prime}\rangle\Big|_{T\neq 0}\nn\\
     &=&\frac{1}{L}\sum\limits_{k}\Big[\frac{E_k}{\epsilon_k}(N_k+1)-1\Big].
  \eeqa

From \Eq{eq}, the $M$ body correlation function at weak coupling regime is
\beq
g_M^w=n^M\Big(1+\frac{M(M-1)}{2\pi n}\int_{-\infty}^{\infty}dk\Big[\frac{E_k}{\epsilon_k}(N_k+1)-1\Big]\Big).
\eeq
\section{Appendix C}
Consider the integral term in \Eq{g_n2}. There are $j$ points between $0$ and $x$ and $(N-j)$ points between $x$ and $L$ among the $(N+2)$ particles. Here we present the calculation of the following  form of the integration:
\beqa
&&\frac{1}{\mathcal{N}^2}\int\ldots\int_{\tilde{x}_1\leq\ldots\leq\tilde{x}_{N+2}} dx_1\ldots dx_N(j+1)\dfrac{\partial}{\partial x}|\psi_{N+2}^F|^2\nn\\
&=&\frac{1}{\mathcal{N}^2}\dfrac{\partial}{\partial x}\sum\limits_{j=0}^{N}\dfrac{(j+1)}{j!(N-j)!}\int_{0}^{x}\ldots\int_0^x dx_1\ldots dx_j\int_{x}^L\ldots\int_x^L dx_{j+1}\ldots,dx_{N}|\psi_{N+2}^F|^2\nn\\
&=&\frac{1}{\mathcal{N}^2}\dfrac{1}{N!}\dfrac{\partial}{\partial x}\dfrac{\partial}{\partial \zeta}\zeta\sum\limits_{j=0}^{N}\zeta^{j}
\left(\begin{array}{c}
N\\
j
\end{array}
\right)
\int_{0}^{x}\ldots\int_0^x dx_1\ldots dx_j\int_{x}^L \ldots\int_x^L dx_{j+1}\ldots,dx_{N}
|\psi_{N+2}^F|^2\Big|_{\zeta=1}\nn\\
&=&\frac{1}{\mathcal{N}^2}\dfrac{1}{N!}\dfrac{\partial}{\partial x}
\Bigg[\dfrac{\partial}{\partial \zeta}\zeta
\prod\limits_{l=1}^N\Big( \int_{x}^{L}+\zeta\int_0^{x}\Big)dx_l
|\psi_{N+2}^F|^2\Bigg]\Bigg|_{\zeta=1}\nn\\
&=&\frac{1}{\mathcal{N}^2}\dfrac{1}{ N!}\dfrac{\partial}{\partial x^F}
\Bigg[\dfrac{\partial}{\partial \zeta}\zeta
\prod\limits_{l=1}^N\Big(\Big[ \int_{0}^{L}- \int_0^{x}\Big]+\zeta\int_0^{x}\Big)dx_l
|\psi_{N+2}^F|^2\Bigg]\Bigg|_{\zeta=1}\nn\\
&=&\frac{1}{\mathcal{N}^2}\dfrac{1}{N!}\dfrac{\partial}{\partial x}
\Bigg[\dfrac{\partial}{\partial \zeta}\zeta
\prod\limits_{l=1}^N\Big( \int_{0}^{L}+(\zeta-1)\int_0^{x}\Big)dx_l
|\psi_{N+2}^F|^2\Bigg]\Bigg|_{\zeta=1}\nn\\
&=&\frac{1}{\mathcal{N}^2}\dfrac{1}{N!}\dfrac{\partial}{\partial x}
\Bigg[
\prod\limits_{l=1}^N \int_{0}^{L} dx_l
|\psi_{N+2}^F|^2+\dfrac{\partial}{\partial \zeta}\zeta\sum\limits_{l=0}^{N}\int_0^{x}(\zeta-1) dx_l
\prod\limits_{j=1,j\neq l}^N \int_0^L  dx_j
|\psi_{N+2}^F|^2\Bigg]\Bigg|_{\zeta=1}+O((\zeta-1)^2)\nn\\
&=&\frac{1}{\mathcal{N}^2}\dfrac{1}{N!}\dfrac{\partial}{\partial x}
\Bigg[
\prod\limits_{l=1}^N \int_{0}^{L} dx_l
|\psi_{N+2}^F|^2+N\dfrac{\partial}{\partial \zeta}\zeta(\zeta-1)\int_0^{x} dx_l
 \prod\limits_{j=1,j\neq l}^N \int_0^L dx_j
|\psi_{N+2}^F|^2\Bigg]\Bigg|_{\zeta=1}\nn\\
&=&\frac{1}{\mathcal{N}^2}\dfrac{1}{N!}\dfrac{\partial}{\partial x}
 \prod\limits_{l=1}^N\int_{0}^{L} dx_l^F
|\psi_{N+2}^F|^2 +
\dfrac{\partial}{\partial x}\int_0^{x} dx_l
\frac{1}{\mathcal{N}^2}\dfrac{1}{(N-1)!}\prod\limits_{j=1,j\neq l}^N \int_0^L dx_j
|\psi_{N+2}^F|^2\nn\\
&=&\alpha^{N+1}\Big(\dfrac{\partial}{\partial x}g_{2}^{N+2,F}(0,x)+\dfrac{\partial}{\partial x}\int_0^{x} dx_l\;g_{3}^{N+2,F}(0,x,x_l)\Big),\label{gn_2}
\eeqa
where we can write
\beqa
\prod\limits_{l=1}^{N}\Big( \int_{x}^{L}+\zeta\int_0^x\Big)dx_l
|\psi_{N}^F|^2
&=&\sum\limits_{j=0}^{N}\zeta^{j}
\left(\begin{array}{c}
N\\
j
\end{array}
\right)
\prod\limits_{l=1}^{N} \int_{0}^{L}dx_l
|\psi_{N}^F|^2.
\eeqa

What below gives a further demonstration on spacial cases, see \Eq{cal1} and \Eq{cal2}:
 Considering a system with $N=3$,
\beqa
&&\prod\limits_{l=1}^3\Big( \int_{x}^{L}+\zeta\int_0^x\Big)dx_l
|\psi_3^F|^2\nn\\
&=&\Bigg[ \iiint_{x}^{L}dx_1dx_2dx_3+\Big(\iint_x^Ldx_1dx_3\;\zeta\int_0^xdx_2 +\iint_x^Ldx_2dx_3\;\zeta\int_0^xdx_1+ \iint_{x}^{L}dx_1dx_2\;\zeta\int_0^xdx_3\Big)\nn\\
&&\quad +\Big(\int_x^Ldx_3\zeta^2\iint_0^xdx_1dx_2+\int_x^Ldx_1\; \zeta^2\iint_0^xdx_2dx_3+\int_x^Ldx_2\;\zeta^2\iint_0^xdx_1dx_3\Big)
+\zeta^3\iiint_0^xdx_1dx_2dx_3\Bigg] |\psi_3^F|^2,\nn\\\label{cal1}
\eeqa

and
\beqa
&&\sum\limits_{j=0}^{3}\zeta^{j}
\left(\begin{array}{c}
3\\
j
\end{array}
\right)
\prod\limits_{l=1}^3 \int_{0,x}^{x,L}dx_l
|\psi_3^F|^2\nn\\
&=&\Bigg[
 \iiint_x^Ldx_ldx_2dx_3+
 \Big(\zeta\int_0^xdx_1\iint_x^Ldx_2dx_3 +
\zeta\int_0^xdx_2\iint_x^Ldx_1dx_3+
\zeta\int_0^xdx_3\iint_x^Ldx_1dx_2\Big)\nn\\
&&\quad+\Big(\zeta^2\iint_0^xdx_1dx_2\int_x^Ldx_3+
\zeta^2\iint_0^xdx_1dx_3\int_x^Ldx_2+
\zeta^2\iint_0^xdx_2dx_3\int_x^Ldx_1\Big)
+ \zeta^3\iiint_0^xdx_1dx_2dx_3\Bigg]
 |\psi_3^F|^2.\nn\\\label{cal2}
\eeqa

\end{document}